\documentclass[12pt,letter]{article}
\pdfoutput=1
\usepackage{graphicx, epsfig, color,cite}
\usepackage{amsmath}
\usepackage{amssymb}
\usepackage{float}
\usepackage{hyperref}
\usepackage{subcaption}

\def\eslt{\not\!\!\!{E_T}}
\def\to{\rightarrow}

\def\bi{\begin{itemize}}
\def\ei{\end{itemize}}

\def\tchi{\tilde\chi}

\def\tst{\tilde t}

\def\tq{\tilde q}
\def\alt{\lesssim}
\def\agt{\gtrsim}
\def\be{\begin{equation}}  
\def\ee{\end{equation}}  
\def\bea{\begin{eqnarray}}  
\def\eea{\end{eqnarray}}

\begin{document}
\begin{titlepage}
\begin{flushright}
OU-HEP-250204
\end{flushright}

\vspace{0.5cm}
\begin{center}
  {\Large \bf All axion dark matter from supersymmetric models}
\vspace{1.2cm} \renewcommand{\thefootnote}{\fnsymbol{footnote}}

{\large Howard Baer$^{1}$\footnote[1]{Email: baer@ou.edu },
Vernon Barger$^2$\footnote[2]{Email: barger@pheno.wisc.edu},
Dibyashree Sengupta$^{3,4}$\footnote[7]{Email: Dibyashree.Sengupta@roma1.infn.it}\\ and Kairui Zhang$^1$\footnote[5]{Email: kzhang25@ou.edu}
}\\ 
\vspace{1.2cm} \renewcommand{\thefootnote}{\arabic{footnote}}
{\it 
$^1$Homer L. Dodge Department of Physics and Astronomy,
University of Oklahoma, Norman, OK 73019, USA \\[3pt]
}
{\it 
$^2$Department of Physics,
University of Wisconsin, Madison, WI 53706 USA \\[3pt]
}
{\it
$^3$ INFN, Laboratori Nazionali di Frascati,
Via E. Fermi 54, 00044 Frascati (RM), Italy} \\[3pt]
{\it
$^4$ INFN, Sezione di Roma, c/o Dipartimento di Fisica, Sapienza Università di Roma, Piazzale Aldo Moro 2, I-00185 Rome, Italy}

\end{center}

\vspace{0.5cm}
\begin{abstract}
\noindent

Supersymmetric models accompanied by certain anomaly-free discrete
$R$-symmetries $\mathbb{Z}_n^R$ are attractive in that 1. the $R$-symmetry
(which can arise from compactified string theory as a remnant of the
broken 10-d Lorentz symmetry) forbids unwanted
superpotential terms while allowing for the generation of an
accidental, approximate global $U(1)_{PQ}$ symmetry needed to solve the
strong CP problem and 2. they provide a raison d'etre for an otherwise ad-hoc
$R$-parity conservation.
We augment the minimal supersymmetric Standard Model (MSSM)
by two additional $\mathbb{Z}_n^R$- and PQ-charged fields $X$ and $Y$
wherein SUSY breaking at an intermediate scale $m_{hidden}$ leads to
PQ breaking at a scale $f_a\sim 10^{11}$ GeV
leading to a SUSY DFSZ axion.
The same SUSY breaking can trigger $R$-parity breaking via higher-dimensional
operators leading to tiny $R$-violating couplings of order $(f_a/m_P)^N$
so that {\it both} $U(1)_{PQ}$ and $R$-parity emerge as accidental, approximate
symmetries.
For $\mathbb{Z}_4^R$ and $\mathbb{Z}_8^R$, we find only an $N=1$ suppression.
Then the lightest SUSY particle (LSP) of the MSSM becomes unstable
with a lifetime of order $10^{-3}-10^1$ seconds so the LSPs
all decay away before the present epoch.
That leaves a universe with all axion cold dark matter and no WIMPs
in accord with recent LZ(2024) WIMP search results.

\end{abstract}
\end{titlepage}

\section{Introduction}
\label{sec:intro}

One of the successes of the Standard Model (SM) is that it provides a rationale
for why globally conserved quantum numbers like baryon (B) and lepton (L)
number are conserved: they are expected to be accidental, approximate
symmetries that arise only because SM gauge invariance doesn't allow
such terms. They are expected to arise via higher dimension operators\cite{Weinberg:1979sa}
such as the Weinberg dimension-5 operator ${\cal L}\ni (\lambda /m_P)LHLH$,
where $L$ and $H$ are the $SU(2)_L$ lepton and Higgs doublets.
The Weinberg operator gives rise to massive neutrinos as required by
neutrino oscillations.
Higher dimensional operators giving rise to for instance proton decay
are also expected.
In fact, all higher order operators are expected to occur
in the SM effective field theory (EFT), suppressed by appropriate powers of
the Planck mass $m_P$, unless explicitly forbidden.

Of course, the SM is plagued by the well-known finetuning problems known
as the gauge hierarchy problem (GHP) and the strong CP problem.
We will assume here the supersymmetric solution to the GHP as exemplified
in the Minimal Supersymmetric Standard Model (MSSM)\cite{Baer:2006rs} where all
quadratic divergences to the Higgs mass cancel due to the (super)symmetry.
The MSSM is also supported by a variety of virtual effects including
gauge coupling unification\cite{Dimopoulos:1981yj},
radiative electroweak symmetry breaking (REWSB) for a top-quark mass
$m_t\sim 100-200$ GeV\cite{Ibanez:1982fr},
matching theory expectations to the measured value of the Higgs
mass $m_h\simeq 125$ GeV\cite{Carena:2002es} and
precision electroweak\cite{Heinemeyer:2006px}.
We will also assume the axionic solution to the strong CP problem as
exemplified by the SUSY DFSZ axion model\cite{Dine:1981rt,Zhitnitsky:1980tq}
since both the MSSM and DFSZ require
two Higgs doublets and the SUSY DFSZ model provides a Kim-Nilles
solution\cite{Kim:1983dt} to the SUSY $\mu$ problem\cite{Bae:2019dgg}.

Even within the SUSY DFSZ-augmented MSSM (referred to here as the PQMSSM),
a variety of additional issues can occur.
The first of these is the Little Hierarchy Problem (LHP) which has been
exacerbated by lack of SUSY signal at LHC\cite{Baer:2020kwz}.
We believe that part of the LHP arose due to early overestimates of finetuning
in supersymmetric models\cite{Baer:2013gva}.
Using the model-independent measure $\Delta_{EW}$\cite{Baer:2012up},
then many previously popular models such as CMSSM\cite{Kane:1993td},
mGMSB\cite{Dine:1995ag} and mAMSB\cite{Giudice:1998xp} and
varieties of split SUSY\cite{Arkani-Hamed:2004ymt} do indeed have a LHP\cite{Baer:2022dfc} but other models such as NUHM\cite{Matalliotakis:1994ft,Ellis:2002wv,Ellis:2002iu,Baer:2005bu}, nAMSB\cite{Baer:2018hwa} and
natural generalized mirage mediation (nGMM)\cite{Baer:2016hfa}
have large portions of parameter space which
remain natural and so do not suffer from the LHP.
These natural SUSY models are typified by
light higgsinos with mass $\alt 350$ GeV and highly mixed top-squarks
with mass $m_{\tst_1}\alt 2-3$ TeV (due to a large trilinear soft $A$-term,
as is generic in gravity-mediation, and which also lifts $m_h\to 125$ GeV).

A second issue with the MSSM is that it generically allows for large L- and B-
violating processes via the superpotential terms
\be
W_{RPV}\ni \lambda_{ijk}L_iL_jE_k^c+\lambda_{ijk}^\prime L_iQ_jD_k^c
+\lambda_{ijk}^{\prime\prime} U_i^cD_j^cD_k^c +\mu_i^\prime L_iH_u
\ee
(where $i,\ j$ and $k$ are generation indices)
which then requires the rather ad-hoc imposition
of $R$-parity conservation (RPC)\cite{Barger:1989rk,Dreiner:1997uz,Bhattacharyya:1997vv}: $R=(-1)^{3(B-L)-2s}$ (where $s$ is spin).
There are rather severe limits on products of $R$-parity violating (RPV)
couplings due to proton decay
\be
\lambda_{11k}^\prime\lambda_{11k}^{\prime\prime}\alt 10^{-25}
\label{eq:pdecay}
\ee
for $m_{\tq}\sim 1$ TeV and so this motivates the ad-hoc
assumption of RPC that sets these to zero.
Other strong limits from $n-\bar{n}$ oscillation and
double nucleon decay provide bounds of order $\lambda^{\prime\prime}\alt 10^{-3}-10^{-4}$\cite{Goity:1994dq} (depending on sfermion mass).
For more RPV coupling bounds from other processes,
see {\it e.g.} Ref's \cite{Barger:1989rk,Dreiner:1997uz,Bhattacharyya:1997vv}.

An implication of RPC is that the lightest SUSY particle
is absolutely stable and if neutral (like the lightest neutralino $\tchi_1^0$),
then it may provide a good candidate for cold dark matter (CDM)
in the universe\cite{Goldberg:1983nd,Ellis:1983ew}.
For natural SUSY models with light higgsinos, then the thermally-produced (TP)
neutralino relic density is underproduced\cite{Baer:2013vpa} with
$\Omega_{\tchi_1^0}^{TP}h^2\sim 0.01$.
This is not of great concern within the PQMSSM because the axion is expected
to make up the bulk of CDM\cite{Bae:2013bva}.
Also, the reduced abundance of higgsino-like
lightest SUSY particles (LSPs) helped PQMSSM models avoid constraints from
WIMP direct detection experiments since the light higgsinos would make up
typically only $\sim 10\%$ of the CDM\cite{Baer:2016ucr}.
However, under new strong limits
from the LZ experiment in 2024\cite{LZ:2024zvo}, even natural SUSY with a
reduced abundance of TP LSPs seems ruled out\cite{Baer:2016ucr,Martin:2024ytt}\footnote{It may be possible to avoid
  the LZ constraint in the PQMSSM in regions of model parameter space where
  large entropy dilution occurs, when the saxion mass $m_s\alt 2m_{LSP}$ and
when $f_a$ is large $\agt 10^{14}$ GeV\cite{Bae:2014rfa}.} 

A third set of issues pertain to the PQMSSM: 1. from whence does the required
global $U(1)_{PQ}$ arise, 2. how does the (cosmologically favored) PQ scale
$f_a\sim 10^{11}$ GeV arise and 3. does the DFSZ axion have the {\it axion quality}\cite{Kamionkowski:1992mf}
required to satisfy the strong CP problem where an axion
misalignment angle $\bar{\theta}\alt 10^{-10}$ is required\cite{Kim:2008hd}?

A solution to the PQMSSM problems was proposed in
Ref. \cite{Lee:2011dya,Baer:2018avn}
where a discrete $R$-symmetry ${\mathbb Z}_n^R$ was invoked.
Discrete $R$ symmetries are expected to arise from various string
compactifications as the remnant of the breakdown of 10-d Lorentz
symmetry\cite{Kappl:2010yu,Nilles:2017heg}.
The anomaly-free discrete $R$ symmetries that forbid the SUSY $\mu$ term
(the first step in solving the SUSY $mu$ problem), suppress $p$-decay and are
consistent with grand unified reps were tabulated by Lee {\it et al.} in
Ref. \cite{Lee:2011dya} and found to be ${\mathbb Z}_4^R$, ${\mathbb Z}_6^R$,
${\mathbb Z}_8^R$, ${\mathbb Z}_{12}^R$ and ${\mathbb Z}_{24}^R$.
Under $R$-symmetries, the superspace co-ordinates $\theta_a$ transform
non-trivially so that the different elements of superfields transform
differently.
Since the superpotential $W$ contributes to the Lagrangian as
${\cal L}\sim \int d^2\theta W$, then the
superpotential carries an overall $R$-charge $+2$.
The various $R$-charges of MSSM superfields are listed in Table \ref{tab:R}
for the different ${\mathbb Z}_n^R$ symmetries.
\begin{table}[!htb]
\renewcommand{\arraystretch}{1.2}
\begin{center}
\begin{tabular}{c|ccccc}
multiplet & $\mathbb{Z}_{4}^R$ & $\mathbb{Z}_{6}^R$ & $\mathbb{Z}_{8}^R$ & $\mathbb{ Z}_{12}^R$ & $\mathbb{Z}_{24}^R$ \\
\hline
$H_u$ & 0  & 4  & 0 & 4 & 16 \\
$H_d$ & 0  & 0  & 4 & 0 & 12 \\
$Q$   & 1  & 5 & 1 & 5  & 5 \\
$U^c$ & 1  & 5 & 1 & 5  & 5 \\
$E^c$ & 1  & 5 & 1 & 5  & 5 \\
$L$   & 1  & 3 & 5 & 9  & 9 \\
$D^c$ & 1  & 3 & 5 & 9  & 9 \\
$N^c$ & 1  & 1 & 5 & 1  & 1 \\
\hline
\end{tabular}
\caption{Derived MSSM field $R$ charge assignments for various anomaly-free 
discrete $\mathbb{Z}_{N}^R$ symmetries which are consistent with $SU(5)$ or 
$SO(10)$ unification (from Lee {\it et al.} Ref.~\cite{Lee:2011dya}).
}
\label{tab:R}
\end{center}
\end{table}

It was emphasized in Ref. \cite{Lee:2010gv} that the hypothesized
${\mathbb Z}_n^R$ 
could generate the global $U(1)_{PQ}$ required by axions and also forbid
any RPV terms in the superpotential.
It also allows for a Majorana singlet neutrino superfield mass term
(thus allowing for see-saw neutrinos) and it forbids dangerous
dimension-5 operators that could destabilize the proton.

In Ref. \cite{Baer:2018avn}, in a PQMSSM model which invoked two additional
PQ fields $X$ and $Y$, the strongest of the discrete $R$-symmetries,
${\mathbb Z}_{24}^R$, was found to forbid superpotential terms up through
$(1/m_P)^6$, thus allowing for a solution to the axion quality
problem.\footnote{Ref. \cite{Bhattiprolu:2021rrj} emphasizes that for lower
  values of $f_a\sim 10^9$ GeV, then lower orders of $n\sim 8$ or 12 for
  $\mathbb{Z}_n^R$ may be sufficient to solve the axion quality problem.}
Under SUSY breaking via a large trilinear soft term, then the $X$ and $Y$
fields developed vevs of order $\sim 10^{11}$ GeV, thus spontaneously
breaking the global PQ symmetry and generating an axion decay constant
$f_a\sim 10^{11}$ GeV, in the cosmologically favored sweet spot where the axion
provides the bulk of the CDM relic density.
The SUSY $\mu$ term was also generated with
$\mu =\lambda_\mu f_a^2/m_P\sim m_{weak}$.

We begin with the PQMSSM model of Ref. \cite{Baer:2018avn}
(base model $B_{II}$ of \cite{Bhattiprolu:2021rrj}) with superpotential
\bea
W &=& f_u QH_u U^c+f_d QH_d D^c +f_\ell L H_d E^c+f_\nu L H_u N^c\\
&+& f X^3Y/m_P+\lambda_\mu X^2 H_u H_d/m_P+ M_N N^c N^c/2
\eea
where generation indices are suppresssed.
We focus on the PQ portion of the scalar potential where
intermediate-scale field vevs for $X$ and $Y$ dominate over the EW-scale vevs
that also develop via REWSB as usual.
Then the $F$-term scalar potential is given by
\be
V_F= |3f \phi_X^2\phi_Y/m_P|^2+|f\phi_X^3/m_P|^2
\ee
and is augmented by
\be
V_{soft}\ni m_X^2|\phi_X|^2+m_Y^2|\phi_Y|^2+(f A_f\phi_X^3\phi_Y/m_P+h.c.)
\ee
where the $\phi_{X,Y}$ are the scalar components of the $X$ and $Y$ PQ
superfields. Minimization of the scalar potential yields vevs
$v_X,\ v_Y\sim 10^{11}$ GeV thus breaking the $\mathbb{Z}_n^R$ and global
$U(1)_{PQ}$ symmetries with axion decay constant $f_a=\sqrt{v_X^2+9v_Y^2}$ and
$\mu =\lambda_\mu v_X^2/m_P$.
For the special case of an assumed $\mathbb{Z}_{24}^R$ symmetry, then
all non-renormalizable operators suppressed by powers of $(1/m_P)^6$
or less are forbidden thus solving the axion quality problem.

However, now we concern ourselves with non-renormalizable operators
containing $R$-parity violating combinations:\footnote{Our methods bear
  some similarities to the analysis presented in Ref. \cite{Bar-Shalom:2003jlq}.}
\be
W\ni (X/m_P)^p(Y/m_P)^q \times (LLE^c\ {\rm or}\ LQD^c\ {\rm or}\ U^cD^cD^c)
\ee
where the $\mathbb{Z}_n^R$ charges for the $R$-parity violating combinations
are listed in Table \ref{tab:QQQ} and the superpotential charge is
$Q(W)=2\ mod\ n =2,2+n,2+2n\cdots$.
\begin{table}[!htb]
\renewcommand{\arraystretch}{1.2}
\begin{center}
\begin{tabular}{c|ccccc}
combination & $\mathbb{Z}_{4}^R$ & $\mathbb{Z}_{6}^R$ & $\mathbb{Z}_{8}^R$ & $\mathbb{Z}_{12}^R$ & $\mathbb{Z}_{24}^R$ \\
\hline
$LLE^c$ & 3  & 11  & 11 & 23 & 23 \\
$LQD^c$ & 3  & 11  & 11 & 23 & 23 \\
$U^cD^cD^c$ & 3  & 11  & 11 & 23 & 23 \\
$LH_u$ & 1 & 7 & 5 & 13 & 25\\
\hline
\end{tabular}
\caption{Derived MSSM field $R$ charge assignments for $R$-parity violating
  combinations for various anomaly-free 
discrete $\mathbb{Z}_{n}^R$ symmetries which are consistent with $SU(5)$ or 
$SO(10)$ unification.
}
\label{tab:QQQ}
\end{center}
\end{table}

\section{Trilinear $R$-parity violation from $\mathbb{Z}_n^R$ models}

Here, we will adopt the notation that $QQQ$ stands for a trilinear
product of visible sector superfields which are RPV.

\begin{enumerate}
\item $\mathbb{Z}_4^R$ case:
For the $\mathbb{Z}_4^R$ case, we can take $R$-charges as $Q_X=+1$ and $Q_Y=-1$.
Then a lowest order operator including a $QQQ$ combination is
$W\ni \lambda (Y/m_P)QQQ$ and so when $X$ and $Y$ obtain vevs, we obtain
an $R$-parity violating coupling of order $(v_Y/m_P)\sim 10^{-7}$.

\item $\mathbb{Z}_6^R$ case:
For the $\mathbb{Z}_6^R$ case, we can take $R$-charges as $Q_X=-1$ and $Q_Y=11$.
Then a lowest order operator including a $QQQ$ combination is
$W\ni \lambda (X/m_P)^3QQQ$ and so when $X$ and $Y$ obtain vevs, we obtain
an $R$-parity violating coupling of order $(v_X/m_P)^3\sim 10^{-21}$.

\item $\mathbb{Z}_8^R$ case:
For the $\mathbb{Z}_8^R$ case, we can take $R$-charges as $Q_X=-1$ and $Q_Y=5$.
Then a lowest order operator including a $QQQ$ combination is
$W\ni \lambda (X/m_P)^1QQQ$ and so when $X$ and $Y$ obtain vevs, we obtain
an $R$-parity violating coupling of order $(v_X/m_P)^1\sim 10^{-7}$.

\item $\mathbb{Z}_{12}^R$ case:
For the $\mathbb{Z}_{12}^R$ case, we can take $R$-charges as $Q_X=-1$ and
$Q_Y=5$. Then a lowest order operator including a $QQQ$ combination is
$W\ni \lambda (Y/m_P)^3 QQQ$ and so when $X$ and $Y$ obtain vevs, we obtain
an $R$-parity violating coupling of order $(v_Y/m_P)^3\sim 10^{-21}$.

\item $\mathbb{Z}_{24}^R$ case:
For the $\mathbb{Z}_{24}^R$ case, we can take $R$-charges as $Q_X=-1$ and
$Q_Y=5$. Then a lowest order operator including a $QQQ$ combination is
$W\ni \lambda (X/m_P)^2(Y/m_P) QQQ$ and so when $X$ and $Y$ obtain vevs, we obtain
an $R$-parity violating coupling of order $(v_X^2 v_Y/m_P^3)\sim 10^{-21}$.
\end{enumerate}

Thus, to summarize, under the $\mathbb{Z}_4^R$ and $\mathbb{Z}_8^R$
discrete $R$-symmetries, we expect $R$-parity violating processes with
couplings $\lambda\sim 10^{-7}$ whilst under the
$\mathbb{Z}_6^R$, $\mathbb{Z}_{12}^R$ and  $\mathbb{Z}_{24}^R$
symmetries we expect RPV with $\lambda \sim 10^{-21}$.
For the $n=4$ and 8 cases, thus some additional RPV coupling suppression
of some of the couplings (such as lepton or baryon triality\cite{Ibanez:1991pr,Ibanez:1992ji})
would be required to fulfill the $p$-decay constraint Eq. \ref{eq:pdecay}.
Lepton triality $L_3$ would also forbid the potentially troublesome
bilinear RPV terms from arising.

The expected tiny RPV couplings are small enough that the $\tchi_1^0$ LSP
should be stable on the scale of collider experiments, so the usual
SUSY $\eslt$ signatures\cite{Baer:2020kwz} should ensue.
The rough scale for visibility of LSP decay in LHC collider detectors
is that $d=\beta c\gamma\tau\alt 5$ m.
Approximating this as $d\alt c\tau$, then we expect
$\tau\alt 2\times 10^{-8}$ s. 

But on longer time-scales-- such as those affecting cosmology--
then other constraints become relevant.
Under RPV, the (photino-like) $\tchi_1^0$ decay width for a single RPV coupling
is given by\footnote{Some more complete $\tchi_1^0$ decay formulae are given
in Ref. \cite{Dreiner:1994tj}.}
\be
\Gamma (\tchi_1^0 )= \frac{3\alpha\lambda_{111}^{\prime 2}}{128\pi^2}
\frac{m_{\tchi_1^0}^5}{m_{soft}^4}
\ee
where $m_{soft}$ is the SUSY sfermion mass scale which can be in the $10-40$ TeV
range for natural SUSY under $\Delta_{EW}$ emergent from the
string landscape\cite{Baer:2017uvn}.
Including gaugino and higgsino mixing angle factors can reduce this rate
whilst including more decay modes other than just $\lambda_{111}^\prime \ne 0$
will increase the decay rate-- thus, we take this formula to be 
an order-of-magnitude estimate.

\begin{figure}[htb!]
\centering
    {\includegraphics[height=0.5\textheight]{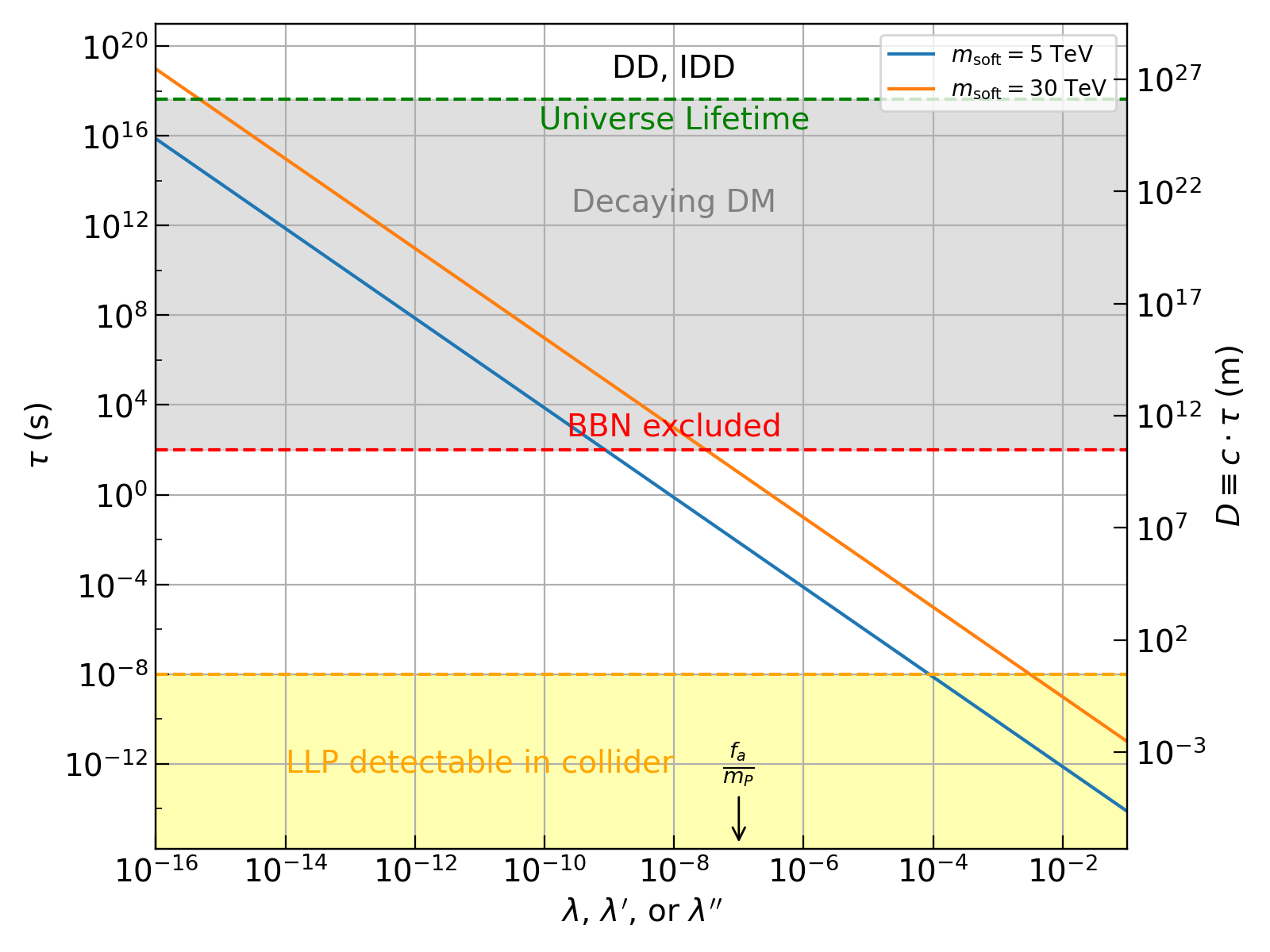}}
    \caption{Lifetime of lightest neutralino (left vertical axis)
      versus RPV couplings
      $\lambda$, $\lambda^\prime$ or $\lambda^{\prime\prime}$ for
      $m_{soft}=5$ and 30 TeV and $m_{\tchi_1^0}=200$ GeV.
      On the right vertical axis, we show the approximate decay
      length D (meters).
                \label{fig:tau_z1}}
\end{figure}

The associated lifetime $\tau_{\tchi_1^0}$ is shown vs. coupling
$\lambda$ in Fig. \ref{fig:tau_z1} for $m_{soft}=5$ and 30 TeV.
The lifetime will obviously shorten if more decay modes are included.

The lifetime of late-decaying neutral relics from the Big Bang is severely
constrained by Big Bang Nucleosynthesis (BBN) which requires the relic
particle decay debris not to affect the successful predictions of light
element abundances as calculated in the standard cosmology.
We adopt the results from Fig. 10 of Jedamzik\cite{Jedamzik:2006xz}
where BBN-allowed regions of late decaying relics is plotted in the
{\it would-be} LSP abundance $\Omega_\chi h^2$ vs. $\tau_\chi$ plane for
relic particle $\chi$.
For our natural SUSY case with a higgsino-like LSP, then
$\Omega_{\tchi_1^0}^{TP}h^2\sim 0.01$ and for $m_{\tchi_1^0}\sim 100$ GeV, then
the lifetime $\tau$ is constrained by BBN to be $\alt 10^2$ s.
This bound is indicated in Fig. \ref{fig:tau_z1} by the red-dashed line,
and the allowed region is below it.
From the plot, we see that
$\lambda$ values $\alt 10^{-8}-10^{-9}$ are then excluded. 
For $\tau \agt \tau_u\sim 4.3\times 10^{17}$ s
($\tau_u$ is the age of the universe) then such $\lambda$ values would
again be possible, but then perhaps excluded by the LZ(2024)
direct WIMP detection (DD) constraints.
Also, for $\tau\agt \tau_u$, then WIMPs would be suspectible to indirect
detection (IDD) via WIMP-WIMP annihilation to gamma rays or antimatter or via
decaying dark matter searches.
In particular, the above $\mathbb{Z}_n^R$ cases with $\lambda\sim 10^{-21}$
may be LZ-excluded whilst the cases $\mathbb{Z}_4^R$ and $\mathbb{Z}_8^R$
with $\lambda\sim 10^{-7}$ are allowed. For these latter cases, the
{\it would-be} relic higgsino-like WIMPs all decay away on times scales
of order $10^{-3}-10^{1}$ s after the Big Bang. Since our considerations
all arise from an assumed SUSY DFSZ axion model, then we may infer that
the coherent-oscillation (CO) produced axions\cite{Abbott:1982af,Preskill:1982cy,Dine:1982ah} comprise all the
remaining dark matter from this class of supersymmetric models.
This seems to be in accord with recent results from the LZ
experiment\cite{LZ:2024zvo} which are strong enough to exclude
thermally-underproduced natural higgsino-like WIMPs,
assuming no nonthermal processes such as entropy-dilution are occurring
(see Fig. 8a of Ref. \cite{Baer:2025zqt}).
We also show in Fig. \ref{fig:tau_z1} the yellow-shaded region wherein
late-decaying LSPs (long-lived particles of LLPs) produced at the LHC
collider might decay within the detector geometry.

\section{Conclusions}
\label{sec:conclude}

The MSSM augmented by a discrete $\mathbb{Z}_n^R$ symmetry is well-motivated
in that the discrete $R$-symmetry forbids the $\mu$ term and other
unwanted superpotential terms while allowing for desired ones.
In an example two-extra field model where
the MSSM is augmented by $X$ and $Y$ fields, then the $\mathbb{Z}_n^R$
symmetry can generate both the global $U(1)_{PQ}$ (as an accidental, approximate
global symmetry) which solves the strong CP problem with
$f_a\sim 10^{11}$ GeV (cosmological sweet spot for axion dark matter)
and generates a Kim-Nilles $\mu$ term $\sim 100-350$ GeV
while also forbidding RPV superpotential terms as renormalizable operators.
The higher $\mathbb{Z}_n^R$ symmetries can also solve the axion quality problem.
RPV couplings can be generated via non-renormalizable operators with
expected magnitude $(f_a/m_P)^N$ where $N=1$ for $\mathbb{Z}_4^R$
and $\mathbb{Z}_8^R$ and $N=3$ for the remaining $n=6,\ 12$ and $24$.
(Thus, both $U(1)_{PQ}$ and $R$-parity emerge from the underlying
$\mathbb{Z}_n^R$ as approximate, accidental symmetries.)
The $N=1$ case leads to suppressed RPV couplings expected of order
$\lambda_{ijk},\lambda^\prime_{ijk}$ and $\lambda_{ijk}^{\prime\prime}\sim 10^{-7}$.
In such a case, the LSP of the MSSM, a thermally-underproduced higgsino-like
WIMP, would be unstable with lifetime of order $\tau\sim 10^{-3}-10$ seconds.
This would result in SUSY models with {\it all} SUSY DFSZ axion dark matter
since the WIMPs will have decayed before BBN is finished. These results are
in accord with recent WIMP search results from LZ(2024)\cite{LZ:2024zvo}
which seem to rule out thermally-produced natural SUSY WIMPs,
even with a diminished abundance. The SUSY DFSZ axions have a very
diminished $a\gamma\gamma$ coupling due to the presence of higgsinos
circulating in the triangle diagram and so are beyond present
axion haloscope detection capabilities\cite{Bae:2017hlp}.

The regions of parameter space we find are in accord with the analysis of
Higaki {\it et al.}\cite{Higaki:2014eda} showing how Affleck-Dine
baryogenesis can occur with tiny RPV couplings $\lambda\sim 10^{-9}-10^{-6}$ and with $m_{3/2}\agt 10$ TeV.
See also Akita and Otsuka\cite{Akita:2018zma}.
See also Ref. \cite{Sikivie:2024ffa} wherein it is argued that axion
dark matter is the reason that supermassive black holes form at
cosmic dawn (at redshift $z\sim 10$).

{\it Acknowledgements:} 
We thank X. Tata for comments on the manuscript.
HB gratefully acknowledges support from the Avenir Foundation.
VB gratefully acknowledges support from the William F. Vilas estate.


\bibliography{axdm2025}
\bibliographystyle{elsarticle-num}

\end{document}